\documentclass[aps,prl,reprint,showpacs,showkeys,preprintnumbers]{revtex4-2}

\usepackage[bookmarks=true]{hyperref}
\usepackage{graphicx,epsfig,color,xcolor}
\usepackage[english]{babel}
\usepackage{amssymb,amsfonts,amsmath}
\usepackage{verbatim}
\usepackage{ulem}

\newcommand{\be}{\begin{equation}} \newcommand{\ee}{\end{equation}}
\newcommand{\bea}{\begin{eqnarray}} \newcommand{\eea}{\end{eqnarray}}

\newcommand{\dd}{\text{d}}

\newcommand{\Msol}{M_{\odot}}

\newcommand{\tH}{t_{\text{\tiny H}}}

\newcommand{\MH}{M_{\text{\tiny H}}}
\newcommand{\RH}{R_{\text{\tiny H}}}
\newcommand{\ene}{\rho} 
\newcommand{\MPBH}{M_{\text{\tiny PBH}}}
\newcommand{\newsec}[1]{\textbf{#1.}}

\begin{document}

\title{Primordial Black Hole Formation during a Strongly Coupled Crossover}

\author{Albert Escrivà}
\email{escriva.manas.albert.y0@a.mail.nagoya-u.ac.jp}
\affiliation{\mbox{Division of Particle and Astrophysical Science, Graduate School of Science,} \\ Nagoya University, Nagoya 464-8602, Japan}
\author{Javier G. Subils}
\email{javier.subils@su.se}
\affiliation{Nordita, Stockholm University and KTH Royal Institute of Technology,\\
Hannes Alfvéns väg 12, SE-106 91 Stockholm, Sweden.}

\begin{abstract}
The final mass distribution of primordial black holes is sensitive to the equation of state of the Universe at the scales accessible by the power spectrum. Motivated by the presence of phase transitions in several beyond the Standard Model theories, some of which are strongly coupled, we analyze the production of primordial black holes during such phase transitions, which we model using the gauge/gravity duality. We focus in the (often regarded as physically uninteresting) case for which the phase transition is just a smooth crossover. We find an enhancement of primordial black hole production in the range $M_{\rm{PBH}}\in[10^{-16},10^{-6}]\Msol$. 
\end{abstract}

\preprint{NORDITA 2022-082}

\pacs{
}

\maketitle

\newsec{Introduction}
Phase transitions (PTs) in the early Universe have received increasing attention since the first gravitational wave (GW) detection \cite{LIGOScientific:2016aoc}. Frequently, the focus is set in first-order PTs, during which  bubbles are nucleated. Their expansion, collision and collapse could lead to a detectable stochastic background of GWs \cite{Hindmarsh:2020hop,Guo:2020grp,Kalogera:2021bya,Caprini:2019egz}. 

In the Standard Model (SM), there is no first-order PT: both deconfinement in Quantum Chromodynamics (QCD) \cite{Aoki:2006we} and electroweak (EW) phase transitions \cite{Kajantie:1996mn, Laine:1998vn, Rummukainen:1998as} are smooth crossovers. These do not lead to bubble nucleation. However, in minimal extensions of the SM \cite{Carena:1996wj,Delepine:1996vn,Laine:1998qk,Huber:2000mg,Grojean:2004xa,Huber:2006ma,Profumo:2007wc,Barger:2007im,Laine:2012jy,Dorsch:2013wja,Damgaard:2015con} the EW PT becomes first-order. First-order PTs also appear in Grand Unified Theories \cite{Georgi:1974sy,Pati:1974yy}. Thus, GWs detections could lead to the discovery of new physics.

In this Letter, we examine the rather ignored but plausible scenario where the theory completing the SM undergoes a smooth crossover (SC), instead of a first-order PT. We show that, despite the absence of bubble formation,  the sudden change in the Equation of State (EoS) of the Universe in such completion of the SM would still have important phenomenological consequences. More precisely, such a phase structure leads to sensitive differences in the abundance of primordial black holes (PBHs) that are expected to be formed.

PBHs are black holes formed in the very early Universe due to the collapse of inflationary cosmological perturbations \cite{1967SvA....10..602Z,Hawking:1971ei,1974MNRAS.168..399C} or other mechanisms \cite{Escriva:2022duf}. Famously, they could constitute all the Dark Matter (DM) or a significant fraction of it \cite{Chapline:1975ojl}. Their abundance is exponentially sensitive to the threshold for PBH formation, related to how big a perturbation has to be to collapse into a black hole. It has been observed that when the pressure of the cosmological fluid decreases from its radiation dominated value, there is an enhancement of PBH production precisely at the scale where such deviation occurs \cite{1974MNRAS.168..399C, 1975ApJ...201....1C,Jedamzik:1996mr}. 
Intuitively, this happens because pressure gradients act against 
gravity, favoring the collapse into black holes of milder perturbations. For instance, this is known to happen during the QCD crossover, where the enhancement is found around $T \approx 178$ MeV, leading to PBHs with masses at the solar mass scale \cite{Jedamzik:1996mr,Byrnes:2018clq,Franciolini:2022tfm,Escriva:2022bwe}. We wish to show the implications of a similar SC being present at energies above the EW scale.
\begin{figure}[t]
\begin{flushright}
\includegraphics[width=0.46\textwidth]{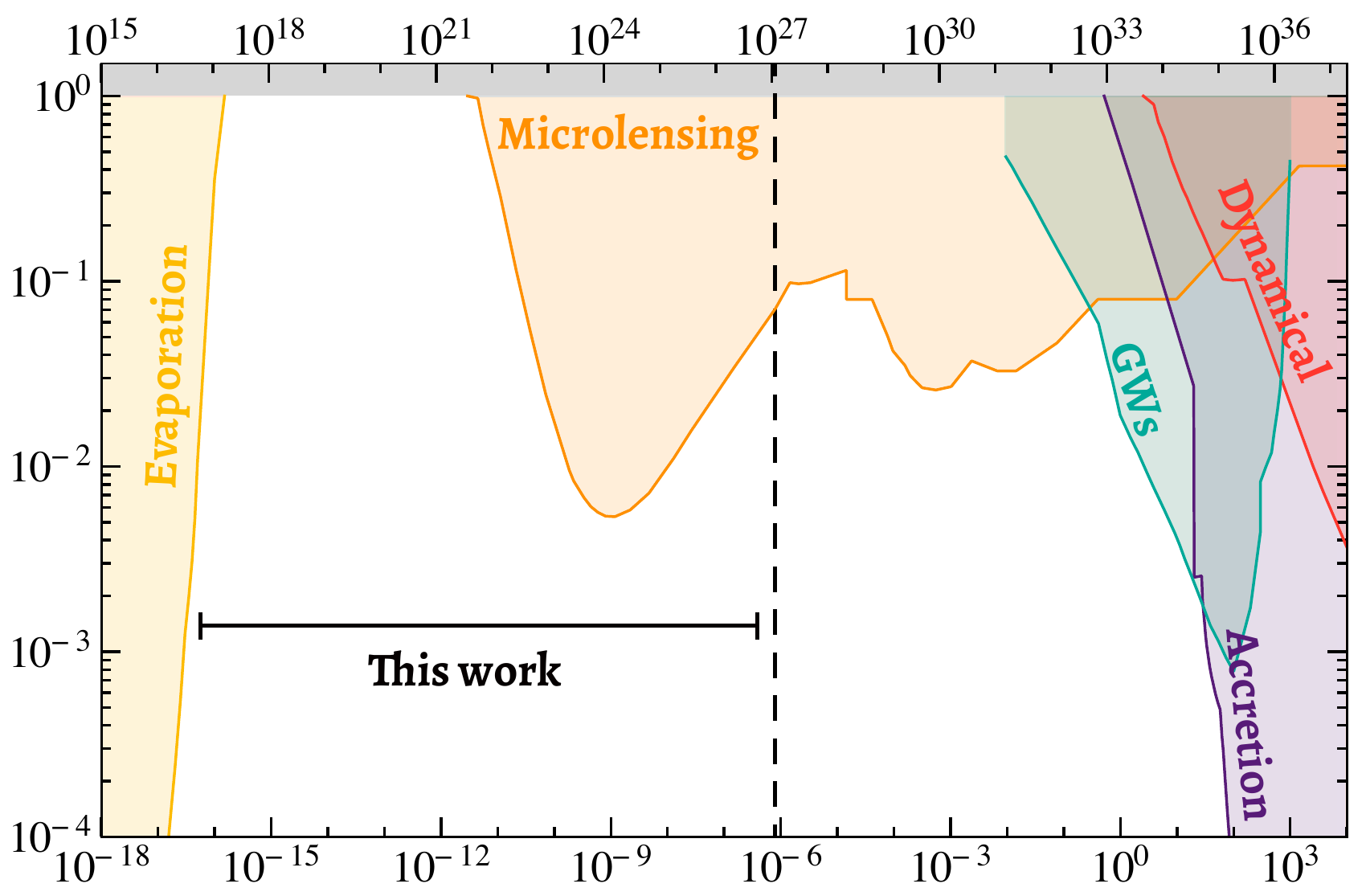}
\put(-245,70){\rotatebox[origin=t]{90}{ $f_{\text{\tiny PBH}}$}}
\put(-130,-10){$\MPBH[\Msol]$}
\put(-130,160){$\MPBH[\rm{g}]$}
\end{flushright}
\caption{Current constraints for the fraction of PBHs, $f_{\text{\tiny PBH}}$, in the form of DM (see Eqs.~\eqref{eq:mass_function},~\eqref{eq:beta} and discussion below them), as a function of the corresponding mass of the PBHs, $\MPBH$. These constraints consider a monochromatic mass function \cite{Carr:2020gox,2021JPhG...48d3001G}. The masses of the PBHs originated from the PT we consider appear in the window spanned by the horizontal black interval. Figure generated using \cite{bradley_j_kavanagh_2019_3538999}.}
 \label{fig:bounds}
\end{figure}

For that, we will be assuming that the theory that completes the SM at high energies is strongly coupled, and we will use the gauge/gravity duality to model its EoS. This assumption is motivated by three reasons. First, models for strongly coupled DM have been considered in the literature \cite{Kribs:2016cew,Tulin:2017ara}, and they potentially lead to the kind of PTs discussed here. Second, the model we will consider has already been used extensively in the literature of bubble nucleation from first-order PTs, for example to compute bubble wall velocities \cite{Bea:2021zsu,Bea:2022mfb} or the expected GW production \cite{Ares:2020lbt,Bea:2021zol}. Finally, when one of the parameters of the model is tuned, the PT becomes a SC. We believe it is relevant to understand what happens in this model when the first-order PT disappears.

Because we are focusing in physics beyond the SM, the PT will take place at a temperature higher than that of the EW scale (around $0.2$ TeV). We will see that the corresponding enhancement appears in the range from $10^{-16}\Msol$ to $10^{-6} \Msol$. Note that this includes the so-called \textit{asteroid mass range}, where no stringent bounds have been found so far \cite{Katz:2018zrn,Montero-Camacho:2019jte}, see Fig.~\ref{fig:bounds}.

\newsec{The model}
The properties of our strongly coupled, beyond the SM fluid will be investigated using the gauge/gravity duality or, in short, holography \cite{Maldacena:1997re}. This correspondence provides a link between states of strongly coupled gauge theories and solutions to classical gravity in one extra dimension. Let us now discuss what these solutions in the gravity side of the duality and the corresponding features of the dual field theory are.

We consider a five-dimensional Einstein-scalar model described by the action
\begin{equation}\label{eq:action}
S\,=\,\frac{1}{16\pi G_5}\int d^5x\sqrt{-g^{[5]}}\left({R}^{[5]} - \frac12 (\partial\phi)^2-V(\phi)\right)\,.
\end{equation}
Here $R^{[5]}$ is the five-dimensional Ricci scalar, $g^{[5]}$ is the determinant of the space-time metric $g^{[5]}_{\mu\nu}$, and $G_5$ is the five-dimensional Newton constant. Note we are working in natural units, $\hbar = c = 1$. Holography relates the scalar field $\phi$ to the gauge coupling of the dual field theory. The fall-off of the scalar $\phi$ near the boundary induces an explicit breaking of conformal invariance. This introduces an energy scale, $\Lambda$, which is related to the critical temperature $T_c$ of the theory.

For simplicity, we assume that the potential $V(\phi)$ comes from a superpotential $W(\phi)$ via the usual relation
\begin{equation}
V(\phi)= -\frac{16}{3}W(\phi)^2 + 8 W'(\phi)^2\,.
\end{equation}
We stress that this choice has nothing to do with supersymmetry. Rather, we choose this particular potential so that the model coincides with that of \cite{Bea:2021zsu,Bea:2022mfb,Ares:2020lbt,Bea:2021zol}, in which the choice is made for convenience. The superpotential reads
\begin{equation}
W(\phi) = -\frac{3}{2}-\frac{\phi^2}{8} -\frac{\phi^4}{64 \phi_M^2} + \frac{\phi^6}{64 \phi_Q}\,.
\end{equation}

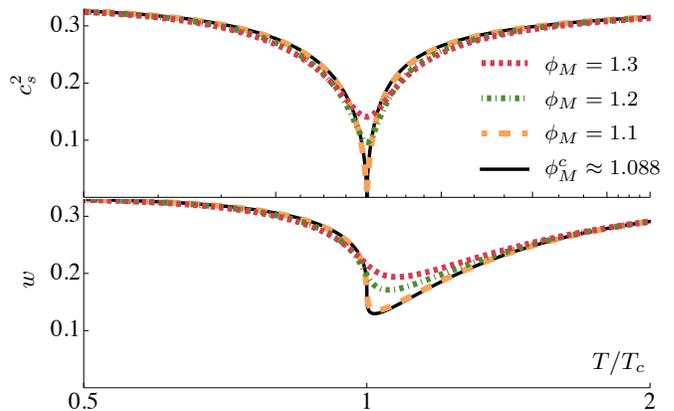
\begin{figure}[t]
\begin{flushright}
\includegraphics[width=0.48\textwidth]{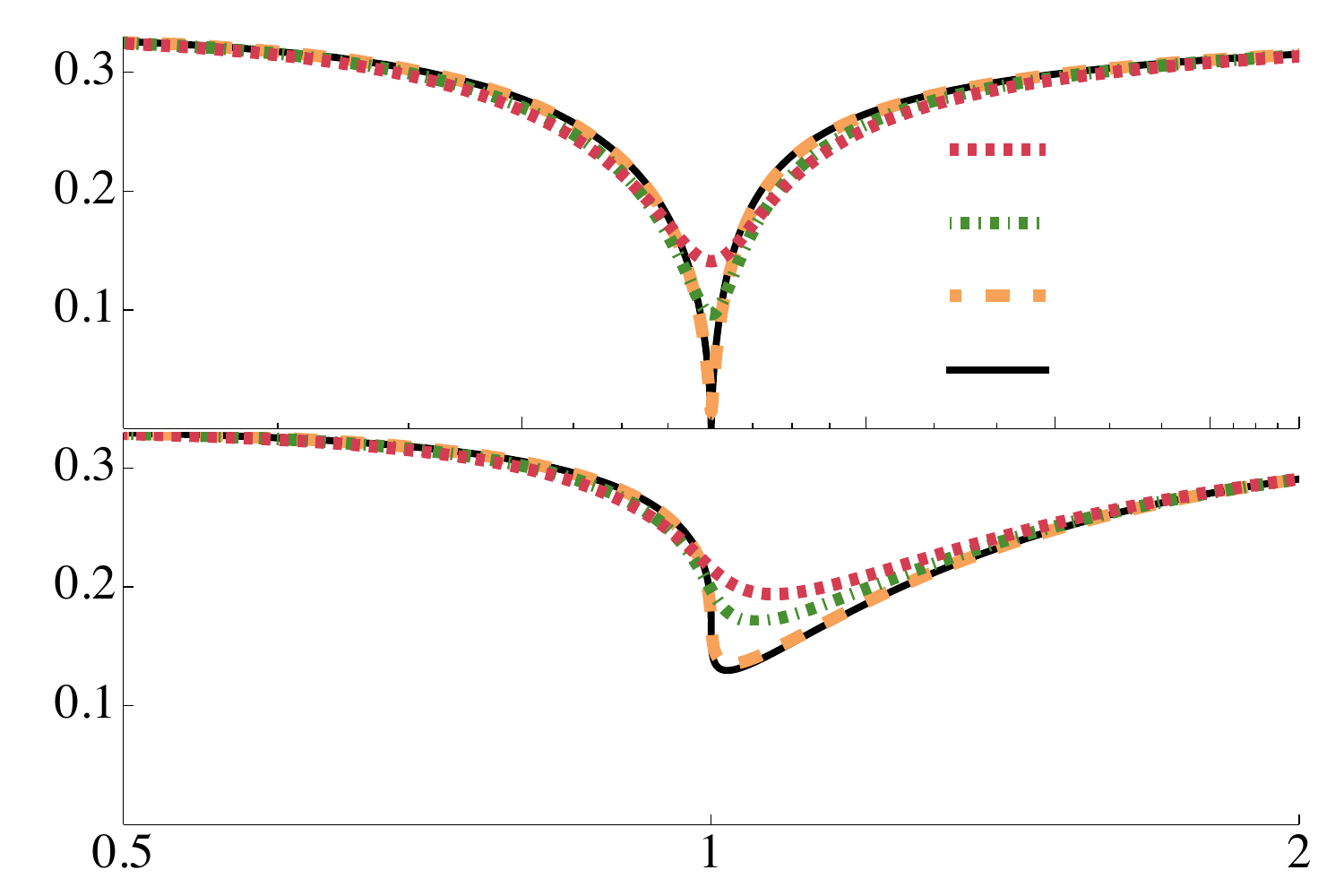}
\put(-30,20){$T/T_c$}
\put(-245,50){\rotatebox[origin=t]{90}{$w$}}
\put(-250,120){\rotatebox[origin=t]{90}{$c_s^2$}}
\put(-48,133){\footnotesize$\phi_M = 1.3$}
\put(-48,120){\footnotesize $\phi_M = 1.2$}
\put(-48,107){\footnotesize $\phi_M = 1.1$}
\put(-48,94){\footnotesize $\phi_M^c \approx 1.088$}
\end{flushright}
\caption{Sound speed squared (\textbf{top}) and EoS (\textbf{bottom}) as a function of the temperature for the different choices of $\phi_M$. When the critical point $\phi_M^c \simeq 1.088$ is reached, the speed of sound vanishes at $T_c$. Both quantities approach $1/3$ at high and low temperatures due to the presence, respectively, of an UV and an IR fixed points.}
\label{fig:w_and_soundspeed}
\end{figure}

The reason why this model has caught so much attention is because by changing the parameters $\phi_M$ and $\phi_Q$ one can easily obtain a first-order PT \cite{Bea:2018whf}. For definiteness, we will fix $\phi_Q = 10$ and let $\phi_M$ vary. Remarkably, the order of the PT changes when $\phi_M$ is tuned: below a certain value $\phi_M < \phi_M^c$ the theory undergoes a first-order PT. When $\phi_M = \phi_M^c$, a critical point where a second-order PT takes place is found. Finally, if $\phi_M \gtrsim \phi_M^c$ there is a SC between the two phases. This is the case we focus on. Numerically, we can determine that $\phi_M^c = 1.088 \pm 0.001$.

We are interested in black brane solutions of \eqref{eq:action} that asymptote to Anti-de Sitter (AdS) space at infinity. We construct them following standard techniques \footnote{See \cite{Gubser:2008ny} for instance. Alternatively, a \textit{shooting method} could be used \cite{Dias:2015nua}, which has the advantage that the energy density and pressure can be computed from the fall off of the energy momentum tensor using holographic renormalization \cite{deHaro:2000vlm}. The latter method gives better control over the numerics}. The properties of these black branes give us the features of the dual plasma. For instance, the temperature $T$ and entropy density $s$ of the states of the plasma are read off from the surface gravity and area density of these black brane solutions, respectively. Next, the pressure $p$ can be obtained by integrating the entropy density,
\begin{equation}
    p(T) = \int_0^T \, s(T') \dd T'\,.
\end{equation}
Finally, the energy density $\ene$ follows from the first law $\ene + p = Ts$. With this information we can construct the two quantities needed to simulate PBH formation, namely the speed of sound squared $c_s^2$ and the ratio between the pressure and the energy density,
\begin{equation}\label{eq:EoS}
    c_s^2 = \frac{\dd p}{\dd \ene } = \frac{s}{T}\frac{\dd T}{\dd s}\,,\qquad w = \frac{p}{\ene }\,.
\end{equation}

Both are shown in Fig.~\ref{fig:w_and_soundspeed} for different choices of $\phi_M<\phi_M^ c$. Note that the critical temperature $T_c$ is defined as the temperature for which $c_s^2$ reaches the minimum.

\newsec{Primordial Black Hole Formation} 
\label{sec:pbh}
So far, we have examined the properties of a strongly coupled fluid, which models a candidate for a completion of the SM. As we have seen, the thermodynamic properties of this fluid depend on a parameter $\phi_M$ that we can adjust, and whose value affects the nature of the PT in its EoS.

The situation that we have in mind is that a such fluid fills the Universe at some point during its cosmological evolution. At this stage, the Universe is approximately homogeneous, isotropic and expanding. Thus, it is well described by a Friedmann–Lema\^itre–Robertson–Walker (FLRW) metric. On top of this background, we find curvature perturbations, seeded, for example, by quantum fluctuations during Inflation. We want to examine when such perturbations collapse into black holes.

For that we will need to solve Einstein's equations,
\begin{equation}\label{eq:Einstein}
    R^{[4]}_{\mu\nu} -\frac{1}{2} g^{[4]}_{\mu\nu}R^{[4]} = 8\pi G_{4} T_{\mu\nu}\,, 
\end{equation}
where $R^{[4]}_{\mu\nu}$ is the Ricci tensor in four dimensions, $R^{[4]}$ stands for its trace, $g^{[4]}_{\mu\nu}$ is the spacetime metric, $G_{4}$ is the four-dimensional Newton's constant and $T_{\mu\nu}$ is the energy-momentum of the fluid. We assume that dissipative effects are not important, meaning that the energy-momentum tensor takes the form of a perfect fluid
\begin{equation}\label{eq:Tmunu_tensor}
    T_{\mu\nu} = (\ene + p) u_\mu u_\nu + p g^{[4]}_{\mu\nu},
\end{equation}
with $u$ the fluid velocity. It is thus given in terms of the energy density and pressure, computed earlier. Moreover, we assume spherical symmetry, which we incorporate into our ansatz for the metric 
\begin{equation}\label{eq:metric}
    \dd s^2 = -A(r,t)^2\dd t^ 2 + B(r,t)^2\dd r^2 + R(r,t)^2 \dd\Omega^ 2\,.
\end{equation}
Here, $\dd\Omega^ 2$ is the metric of a two-sphere with unit radius. Also, we will refer to $t$ as the cosmic time, $A(r,t)$ as the lapse function, and $R(r,t)$ as the areal radius. From the latter we can define the Misner--Sharp mass as the mass inside the surface given by $R(r,t)=$ constant,
\begin{equation}\label{eq:MisnerSharp_Mass}
    M(R) = \int_0^R 4\pi \rho \tilde{R}^ 2 \, \dd\tilde{R}\,.
\end{equation}
With this particular ansatz, the system of Eqs.~\eqref{eq:Einstein} can be expressed in the form worked out by Misner and Sharp \cite{Misner:1964je} in the comoving gauge. For nonconstant EoS like ours, they take the form written in \cite{Escriva:2022bwe}, in which the appropriate initial conditions are also discussed. Like there, we solve them numerically using pseudospectral methods (see also \cite{Escriva:2019nsa}). 

As we anticipated, in our simulations, the Universe is initially described by a very small, superhorizon scale perturbation on top of our fluid at some constant density. If there were no perturbations, the Universe would remain homogeneous and isotropic. We can think of this as the \textit{background} solution to Eqs.~\eqref{eq:Einstein}. In this case, it is customary to use Eq.~\eqref{eq:MisnerSharp_Mass} to define the corresponding horizon mass,
\begin{equation}\label{eq:horizon_mass}
    \MH = \frac{4\pi}{3} \ene_{\rm b} \RH^3 = \frac{1}{4} \sqrt{\frac{3}{2 \pi}} \, \, \left(\ene_{\rm b}G_4^{3}\right)^{-\frac{1}{2}}
\end{equation}
where we have used that $\RH = H^{-1} = (8\pi \ene_{\rm b} G_4 / 3 ) ^{-\frac{1}{2}}$ is the Hubble radius and $\ene_{\rm b}$ depends only on time, with the subindex $\rm b$ standing for ``background.'' It is useful to keep in mind that $G_4^{-3/2}\simeq 1.6  \cdot 10^{-6} \text{TeV}^2 M_{\odot}$ when it comes to expressing our results in solar masses.

On top of this homogeneous Universe, we consider a cosmological perturbation that reenters the horizon at $t = \tH$ (an expression for $\tH$ will be given later). As we will see, there will be an enhancement in the PBH production for perturbations reentering when the SC is taking place. The mass of the statistically significant black holes formed is expected to be comparable to $\MH(\tH)$, the horizon mass evaluated at $\tH$ \cite{1977A&A....56..377C,2019PhRvL.122n1302G}. For that reason we can use Eq.~\eqref{eq:horizon_mass} to estimate the peak in the mass distribution of PBHs. Note that $\ene_{\rm b}$ scales with $T_c^4$ (and is $\rho \simeq 75 T_c^4$ in our model when the transition occurs~\footnote{Note that this means we are fixing $L^3/(8\pi G_5^2)=1$, with $L$
the radius of the asymptotic AdS space.}). This means that $\MH(\tH)$ scales as $T_c^2$. Setting $T_c\simeq 10^{4}$ TeV we would get a peak around $10^{-16}\Msol$, above the constraint coming from Hawking evaporation. Furthermore, we require that $T_c$ is above the temperature of the electroweak PT ($\sim0.2$TeV), below which we trust $\Lambda$CDM-SM. This implies that the position of the peak will be below $10^{-6}\Msol$. Thus, our model produces PBHs in the range of roughly $[10^{-16},10^{-6}]\Msol$ for $T_c\in[0.2, 10^4]$ TeV.

The fluctuations we consider are adiabatic, and will therefore be frozen at superhorizon scales ($t \ll \tH$). Consequently, at these scales the spacetime metric can be modeled by a FLRW metric with a non-constant curvature $K(r)$ \cite{Shibata:1999zs,2007CQGra..24.1405P,2015PhRvD..91h4057H}.
\begin{equation}\label{eq:perturbation}
    \dd s^2 = -\dd t^2 + a(t)^2 \left(\frac{\dd r^2}{1-K(r)r^2}+ r^2 \dd \Omega^2\right)\,.
\end{equation}
This serves to establish the initial conditions to evolve Eqs.~\eqref{eq:Einstein}. Note that $K(r)$ connects to the hydrodynamic variable of the cosmological fluctuation at superhorizon scales \cite{2007CQGra..24.1405P,2015PhRvD..91h4057H}, which is useful to set up the initial conditions for the numerical simulation \cite{Escriva:2022bwe}. Given $K(r)$, the corresponding superhorizon scales, radiation-dominated compaction $\mathcal{C}(r)$ is written as $\mathcal{C}(r) =  (2/3)K(r) r^2$. It represents twice the mass excess with respect to the background solution within the volume of radius $r$. This function generically possesses a maximum at a certain $r=r_m$, which, in turn, sets the amplitude of the cosmological fluctuation, $\delta_m \equiv \mathcal{C}(r_m)$ \cite{Shibata:1999zs,2015PhRvD..91h4057H,2019PhRvD.100l3524M}. 
It also allows us to define the time of horizon crossing as $\RH(\tH) = a(\tH)  r_m$. Gravitational collapse into a black hole will occur for amplitudes above a certain threshold value $\delta_c$, which we discuss next.

\begin{figure}[t]
\begin{flushright}
\includegraphics[width=0.46\textwidth]{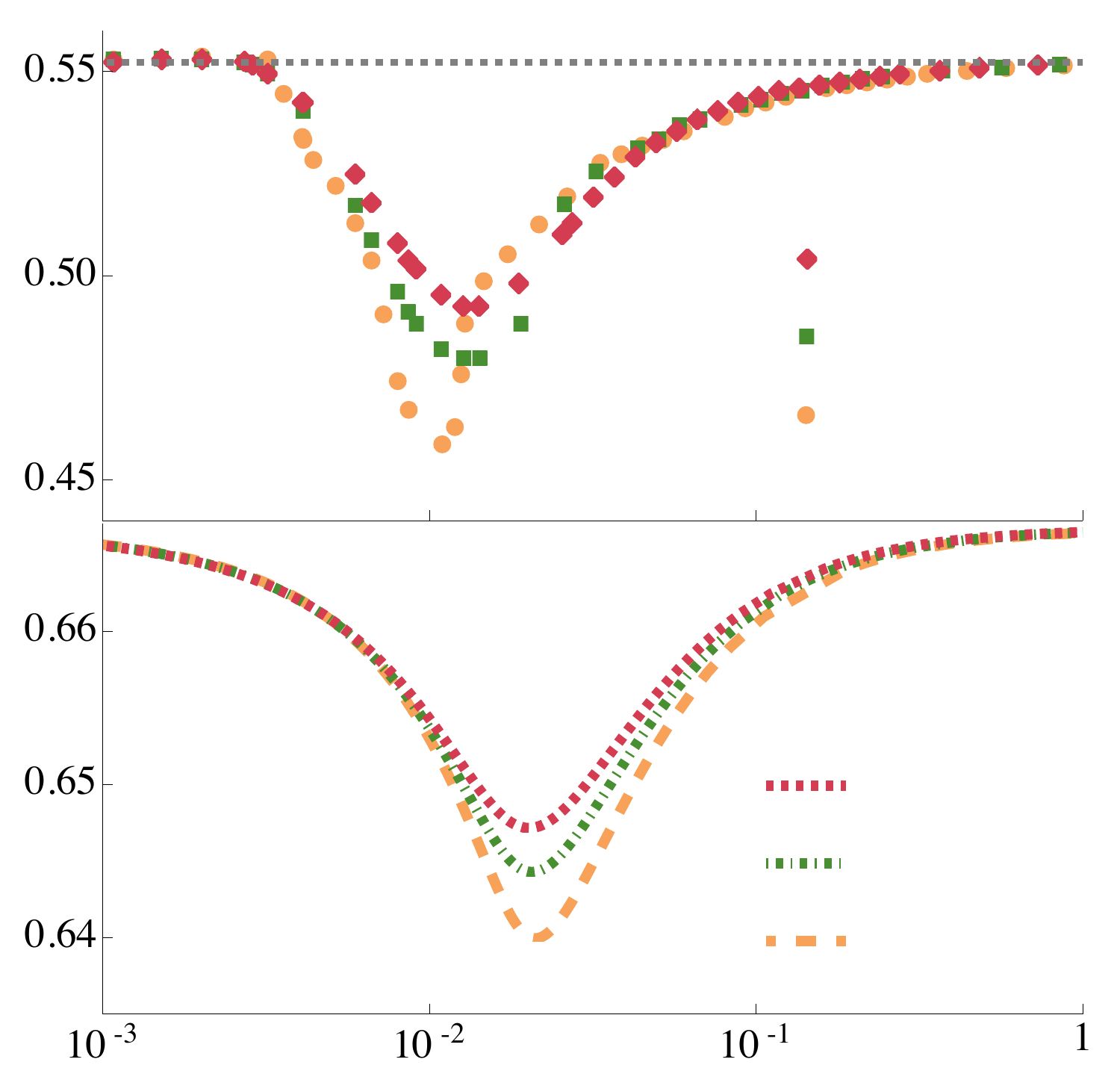}
\put(-135,-10){$\MH(\tH) T_c^2 G_4^{3/2}$}
\put(-245,60){\rotatebox[origin=t]{90}{$\xi_1$}}
\put(-245,160){\rotatebox[origin=t]{90}{$\delta_c$}}
\put(-50,59){$\phi_M = 1.3$}
\put(-50,42.5){$\phi_M = 1.2$}
\put(-50,26){$\phi_M = 1.1$}
\put(-50,169){$\phi_M = 1.3$}
\put(-50,152.5){$\phi_M = 1.2$}
\put(-50,136){$\phi_M = 1.1$}
\end{flushright}
\caption{Threshold values (\textbf{top}) and function $\xi_1$ (\textbf{bottom}) as a function $\MH(\tH)$ for different choices of $\phi_M$. The gray dashed line on the upper panel corresponds to the threshold achieved at the fixed points (where $p= \ene/3$).}
 \label{fig:thresholds}
\end{figure}
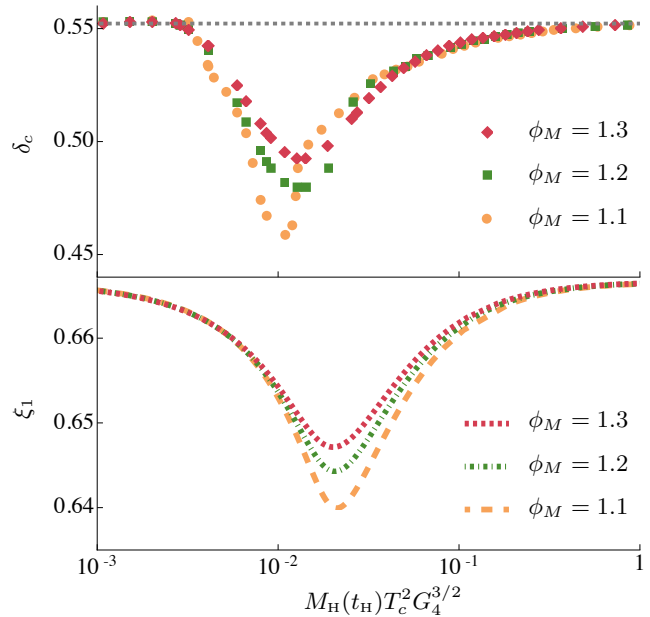

\newsec{Threshold values}
Following \cite{Escriva:2022bwe}, we compute numerically the thresholds of PBH formation for the holographic EoS constructed in Eq.~\eqref{eq:EoS}. Since for our simulations we will take a nearly flat scale-invariant power spectrum (PS) with $n_s \simeq 1$, our perturbations $K(r)$ can be appropriately modeled by a polynomial profile \cite{Escriva:2020tak} with an index $q \approx 3.14$ \cite{2020PhRvD.101d4022E,2021PhRvD.103f3538M}. The results corresponding to this choice are shown in Fig.~\ref{fig:thresholds}. As expected, when $\phi_M$ approaches the critical value, the decrease in $w$ and significant decrease in $c^2_s$ trigger a reduction of the threshold. Remarkably, the diminution in threshold values is much more significant than during the QCD crossover. When taking the same curvature profile $K(r)$ in the latter case, the relative deviation of the threshold with respect to the radiation-dominated era is found to be $\sim  7 \%$ \cite{Escriva:2022bwe} at the peak value. In the present scenario, we encounter sizable reductions near the critical point, of around  $11 \%$, $13 \%$ and $17 \% $ for $\phi_M=1.3,\,1.2$ and $1.1$
respectively.

\newsec{PBHs mass function}
Let us now turn to analyze the phenomenological implications of our smooth beyond the SM PT concerning the distribution of PBHs masses. We start by assuming that sufficiently large primordial density fluctuations leading to PBH formation are Gaussian distributed \cite{Yoo:2020dkz}, with variance $\sigma^2_{\delta \ene}$ and probability density function $P(\delta_m)= \exp\big[-{\delta_m^2}/{(2 \sigma^2_{\delta \ene})}\big]/(2 \pi \sigma^2_{\delta \ene})^{\frac{1}{2}}$.

To obtain concrete quantitative results we are forced to make a choice for the PS. This will be our main assumption, and the final distribution of PBHs masses will depend on this choice strongly. For instance, if we chose a monochromatic PS, the effect of the presence of the crossover would be very mild. In contrast, there are generic qualitative effects present as long as the PS is sufficiently broad and probes all the relevant scales. For concreteness, we consider a nearly flat, scale-invariant PS with shape $\mathcal{P}(k) = \mathcal{A} \,(k/k_{\text{\tiny min}})^{\bar{n}_s-1}\Theta(k-k_{\text{\tiny min}})\Theta(k_{\text{\tiny max}}-k)$. This is a common choice \cite{Carr:2019kxo,DeLuca:2020ioi,DeLuca:2020agl,Sugiyama:2020roc}. The amplitude $\mathcal{A}$ is related to the fraction $f^{\rm tot}_{\text{\tiny PBH}}$ of PBHs that constitute the DM. Additionally, $\bar{n}_s$ is the spectral index at PBH scales, which we take in the range $\bar{n}_s\in[0.955,0.965]$. We stress that our results are also sensitive to this choice but that there are notable generic features that extend for a wider range of $\bar{n}_s$. We will comment on this later. Furthermore, $k_{\text{\tiny max}}$ and $k_{\text{\tiny min}}$ are chosen so that the range of masses $[10^{-16},10^{2}]\, M_{\odot}$ is accessed by the PS. Such a PS is realized in some inflationary models with convenient engineering of the inflationary PS \cite{Wands:1998yp,Leach:2000yw,Leach:2001zf,Byrnes:2018txb,Franciolini:2022pav}.

\begin{figure}[t]
\begin{flushright}
\includegraphics[width=0.45\textwidth]
{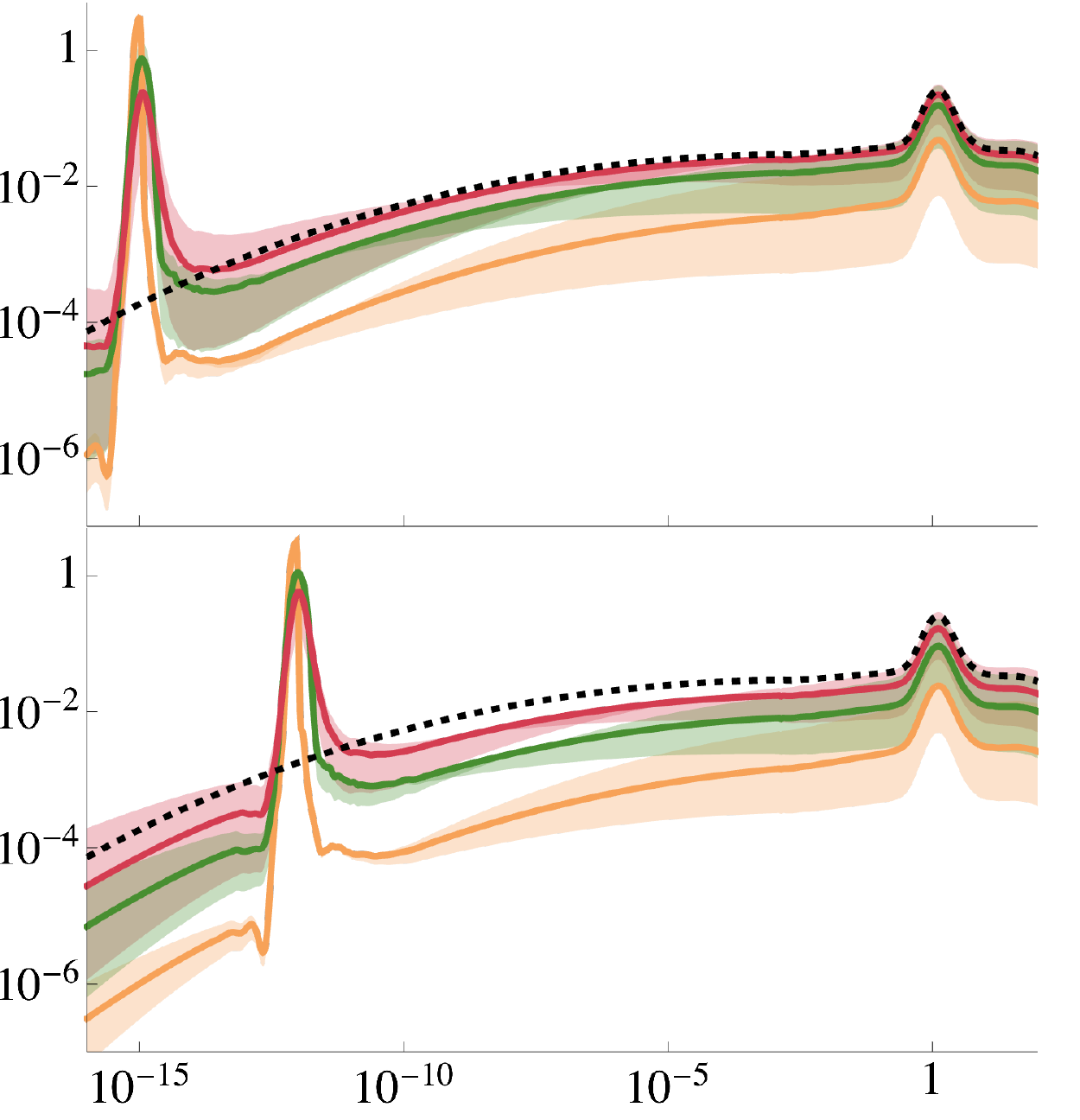}    
\put(-210,3){\includegraphics[width=0.40\textwidth]{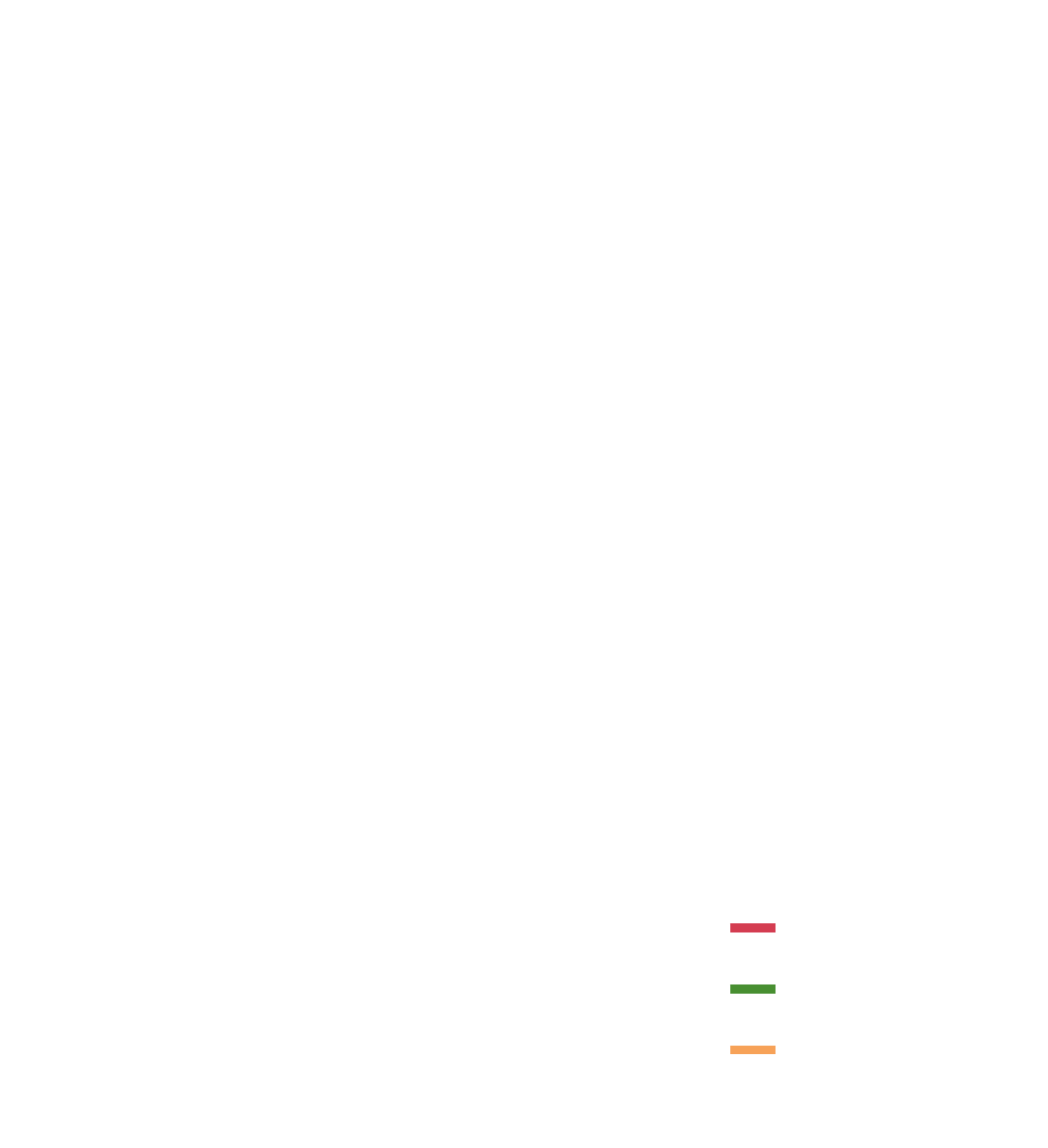}}
\put(-135,20){$\MPBH[\Msol]$}
\put(-130,225){$T_c = 3\cdot 10^3$ TeV}
\put(-130,110){$T_c = 10^2$ TeV}
\put(-54,45){\footnotesize$\phi_M = 1.3$}
\put(-54,32.5){\footnotesize$\phi_M = 1.2$}
\put(-54,20){\footnotesize$\phi_M = 1.1$}
\put(-245,65){\rotatebox[origin=t]{90}{$f(\MPBH)$}}
\put(-245,180){\rotatebox[origin=t]{90}{$f(\MPBH)$}}
\end{flushright}
\caption{\textbf{Top}: PBH mass function computed for a scale-invariant PS and corresponding to $T_c = 3 \cdot 10^3$ TeV. The solid curves correspond to the the choice $\bar{n}_s=0.96$, whereas the bands expands $\bar{n}_s\in[0.955,0.965]$. The case $\bar{n}_s = 0.96$ in the absence of the high-temperature SC is depicted as a black dashed curve for comparison. We have considered $M_{\rm PBH}=\MH(\tH)/2$. \textbf{Bottom}: Same, but with $T_c = 10^2$ TeV.}
 \label{fig:mass_function}
\end{figure}
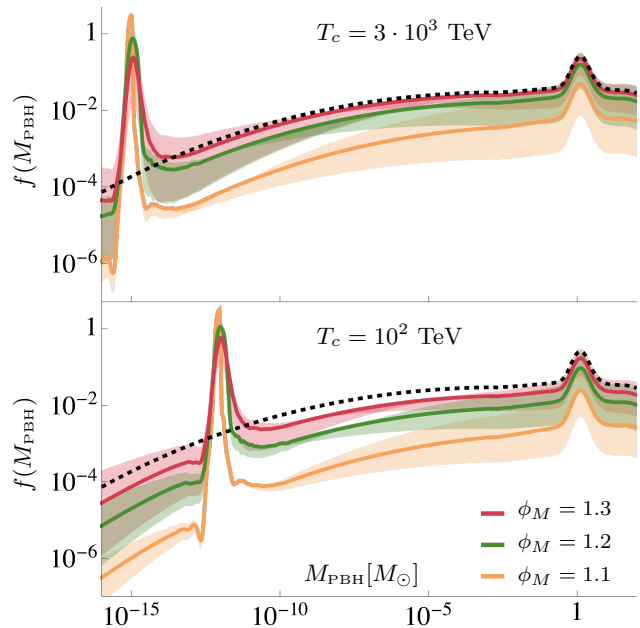

We write the standard deviation of the density perturbations as $\sigma_{\delta \ene}(\MH) = \mathcal{A} \,\xi_{1}(\MH) \MH^{(1-\bar{n}_s)/4}$, where $\xi_1(\MH)$ is a function related to the energy dependence of the EoS \footnote{see Eq.~(2.22) in \cite{Escriva:2022bwe}}  and is shown in Fig.~\ref{fig:mass_function} (\textbf{bottom}). In fact, this gives a good approximation to the actual value of $\sigma_{\delta \ene}(\MH)$. 

Knowing the PS, the relative abundance of PBHs can be computed. For that we will make two approximations. First, we take the mass of the PBH that forms from the collapse of a particular curvature perturbation to be proportional to the horizon mass \eqref{eq:horizon_mass} at its time of horizon crossing, $M_{\text{\tiny PBH}} = \alpha \MH(\tH)$. This is also a common assumption \cite{Carr:2021bzv}. While facilitating computations, it neglects effects near the critical PBH mass regime \cite{1998PhRvL..80.5481N,2013CQGra..30n5009M}. However, these effects are subdominant with respect to the decrease in the threshold value when it comes to estimate PBH abundances. Accounting for these effects, such as a shift in the peak and a change on the shape around the peak of the mass function, is beyond the scope of the present work. Second, we use the Press-Schechter formalism \cite{1974ApJ...187..425P}. Then the mass function, defined so that $f(\MPBH) \dd \ln \MPBH$ is the fraction of PBHs 
between $\MPBH$ and $\MPBH+\dd \ln \MPBH$ \cite{2018CQGra..35f3001S,Byrnes:2018clq}, reads
\begin{equation}
\label{eq:mass_function}
f(\MPBH) = \frac{1}{\Omega_{\text{\tiny CDM}}}\frac{\dd\,  \Omega_{\text{\tiny PBH}}}{\dd \ln \MPBH} = \frac{1}{\Omega_{\text{\tiny CDM}}}\left(  \frac{M_{\rm eq}}{\MPBH} \right)^{1/2} \beta(\MPBH),
\end{equation}
where $\Omega_{\text{\tiny CDM}}= 0.245$ and the PBH mass abundance $\beta$ is
\begin{equation}
\label{eq:beta}
    \beta = 2 \int_{\delta_c}^{\infty} \frac{\MPBH(\delta_m)}{\MH}P(\delta_m) \rm{d}\delta_m= \alpha \, \rm erfc \left( \frac{\delta_c}{\sqrt{2}\sigma_{\delta \ene}} \right)\,.
\end{equation}

In Eq.~\eqref{eq:mass_function}, $M_{\rm eq}=2.8 \cdot 10^{17}\Msol$ corresponds to the horizon mass at the time of matter-radiation equality \cite{Nakama:2016gzw}. The total abundance of PBHs in the form of DM is then $f^{\rm tot}_{\text{\tiny PBH}} = \Omega_{\text{\tiny PBH}} / \Omega_{\text{\tiny CDM}} = \int f(M_{\text{\tiny PBH}}) \dd \ln M_{\text{\tiny PBH}}$. Using the numerical results of the threshold presented in Fig.~\ref{fig:thresholds}~(\textbf{top}), we compute the mass function of PBH formation following Eqs.~\eqref{eq:mass_function} and \eqref{eq:beta}. If all the DM consists of PBHs, $f^{\rm tot}_{\rm PBH}=1$, we get in this scenario that $\mathcal{A}\simeq 10^{-2}$. Therefore, despite the remarkable reduction in the threshold values, we still need a significant enhancement of the amplitude of the PS on the PBH scales (like when only the QCD crossover is considered \cite{Byrnes:2018clq}).

In Fig.\ref{fig:mass_function}, we show the mass function obtained from the holographic model with $T_c =3\cdot10^3$ TeV (\textbf{top}) and $T_c =10^2$ TeV (\textbf{bottom}), together with the QCD crossover. The peak originated during the strongly coupled crossover is patent. As $\phi_M$ approaches $\phi^c_M$ the enhancement strengthens and the peak sharpens. This enhancement causes the rest of the mass function to drop (so that $f^{\rm tot}_{\rm PBH}=1$ in all cases). We have checked that these are generic features for a range of $\bar{n}_s \lesssim 1$ wider than just $\bar{n}_s\in[0.955,0.965]$\footnote{Note this is a similar range to the one used in \cite{Carr:2019kxo}.}, the main difference being the enhanced production of heavy (light) PBHs when $\bar{n}_s$ is sensitively smaller than (closer to) $1$. Actually, it is already clear from Fig.~\ref{fig:mass_function} that a mild modification of the spectral index $\bar{n}_s$ significantly modifies the abundance in the region of the peak at the stellar mass range, corresponding to the QCD crossover.

\newsec{Discussion}
In this Letter, we have studied for the first time implications of the presence of a strongly coupled SC in the very early Universe regarding PBH formation. We saw that it can have a significant impact due to the reduction in the EoS and sound speed of the cosmological fluid, more pronounced the closer the theory flows near the critical point.

A two-peak mass function is obtained as a result with the choice of a particular broad PS. The peaks correspond to the beyond the SM and QCD crossovers. It has been observed that the QCD peak alone seems unable to account for all the DM when inferred merging rates from GW observations are considered \cite{Escriva:2022bwe,Franciolini:2022tfm,Juan:2022mir}. This is not the case in our model, since the presence of our strongly coupled crossover and consequent peak at the asteroid mass range lower the amount of PBHs needed in the solar mass range. Despite its simplicity, the model is flexible enough so that an appropriate choice of $\phi_M$ and $T_c$ may match those inferred merging rates. Better statistical estimation of the PBH abundances would be needed to find accurate results though \cite{Germani:2019zez,Yoo:2020dkz}, as well as better exploration of the critical PBH mass regime.

Our work can be extended in many different directions. First, it is natural to ask what happens when the PT becomes of the first order. There have been already some approaches to this problem \cite{Jedamzik:1999am,Liu:2021svg,Davoudiasl:2019ugw,He:2022amv}. It has an additional complication since, as we mentioned at the beginning, bubbles are expected to nucleate in such scenario. Then PBHs are obtained not only by the collapse of primordial perturbations, but also in the shrinking of false vacuum bubbles in the final stage of the process \cite{Baker:2021nyl,Baker:2021sno,Gross:2021qgx,Kawana:2021tde}. An appropriate approach could be that of \cite{Ecker:2021cvz}, in which holography is used to evolve the stress tensor of the microscopic quantum theory and seed Einstein's equations with its expectation value. Then, the dynamical evolution of the bubble can be performed, without any hydrodynamical approximation, and the interplay between bubble nucleation and PBH formation can be examined.

On the other hand, our results should not depend much on the fact that the theory beyond the SM undergoing a SC is strongly coupled, since the information read off from the holographic model \eqref{eq:action} is just the EoS and the speed of sound; see Eq.~\eqref{eq:EoS}. Thus, it would be interesting to perform similar investigations in weakly coupled models, where we expect to find akin phenomenology. In fact, one could just demand thermodynamic consistency and causality and perform an alike study several EoSs featuring a SC with similar to the ones we discussed \footnote{We thank the anonymous referee for pointing this possible approach out.}.

Additionally, it is well known that the spin of the PBHs formed during a radiation-dominated era is small \cite{Harada:2020pzb} but may be larger for softer EoSs \cite{Harada:2017fjm,Kokubu:2018fxy}. Consequently, the sharp reduction of $c^2_s$ could significantly affect the spin of the PBHs at the corresponding scales. At the same time, stochastic GW background induced by scalar perturbations is also sensitive to $c^2_s$, so it would be desirable to check if for the cases discussed here it lies in the range of LISA frequencies \cite{Bartolo:2018evs,Oncins:2022ydg,Barausse:2020rsu,Cai:2018dig} and constitute an evidence of the presence of such a SC \cite{Abe:2020sqb}.

In summary, we have shown how our SC at energies above the EW scale has significant phenomenological impact. We hope our observations serve as the starting point for exciting future investigations considering PBHs as probes for beyond the SM physics.

\vspace{0.3cm}

\begin{acknowledgments}
{\em  Acknowledgments} We thank Jaume Garriga, Cristiano Germani, David Mateos, Marc Oncins, Mikel Sanchez--Garitaonandia, Yuichiro Tada and Chulmoon Yoo for useful discussions and comments. Nordita is supported in part by NordForsk. A.E acknowledges support from the JSPS Postdoctoral Fellowships for Research in Japan (Graduate School of Sciences, Nagoya University).
\end{acknowledgments}

\appendix

\bibliographystyle{apsrev4-1}

\bibliography{bibfile}

\end{document}